\documentclass[aps,showpacs,preprintnumbers,amsmath,amssymb,superscriptaddress,floatfix,a4paper,nofootinbib,11pt,floatfix,nofootinbib,onecolumn]{revtex4}
\usepackage{graphicx}
\usepackage{bm}
\usepackage{rotating}
\usepackage{array}
\usepackage{xcolor}
\usepackage{amsmath,amssymb}
\usepackage{mathrsfs}
\usepackage{graphicx}
\usepackage{color}
\usepackage{subfigure}
\usepackage{fancyhdr}
\usepackage{multirow}
\usepackage{float}
\usepackage{epsfig}
\usepackage{amsfonts}
\usepackage{bm}

\newcommand{\ba}{\begin{eqnarray}}
\newcommand{\ea}{\end{eqnarray}}

\begin{document}

\title{Deformed Starobinsky model in gravity's rainbow}

\author{Phongpichit Channuie} 
\email{channuie@gmail.com}
\affiliation{College of Graduate Studies, Walailak University, Thasala, Nakhon Si Thammarat, 80160, Thailand}
\affiliation{School of Science, Walailak University, Thasala, \\Nakhon Si Thammarat, 80160, Thailand}
\affiliation{Thailand Center of Excellence in Physics, Ministry of Education, Bangkok 10400, Thailand}

\date{\today}

\begin{abstract}

In the context of gravity's rainbow, we study the deformed Starobinsky model in which the deformations take the form $f(R)\sim R^{2(1-\alpha)}$, with $R$ the Ricci scalar and $\alpha$ a positive parameter. We show that the spectral index of curvature perturbation and the
tensor-to-scalar ratio can be written in terms of $N,\,\lambda$ and $\alpha$, with $N$ being the number of {\it e}-foldings, $\lambda$ a rainbow parameter. We compare the predictions of our models with Planck data. With the sizeable number of {\it e}-foldings and proper choices of parameters, we discover that the predictions of the model are in excellent agreement with the Planck analysis. Interestingly, we obtain the upper limit and the lower limit of a rainbow parameter $\lambda$ and a positive constant $\alpha$, respectively.

\end{abstract}


\maketitle


\section{Introduction}
The prediction of a minimal measurable length in order of Planck length in various theories of quantum gravity restricts the maximum energy that any particle can attain to the Planck energy. This could be implied the modification of linear momentum and also quantum commutation relations and results the modified dispersion relation. Moreover, as an effective theory of gravity, the Einstein general theory of gravity is valid in the low energy (IR) limit, while at very high energy regime (UV) the Einstein theory could in principle be improved.

One of the interesting approaches that naturally
deals with modified dispersion relations is called doubly special relativity \cite{AmelinoCamelia:2000ge,AmelinoCamelia:2000mn,AmelinoCamelia:2003ex}. Then Magueijo and
Smolin \cite{Magueijo:2002xx} generalized this idea by including curvature. The modification of the dispersion relation results by replacing the standard one, i.e. $\epsilon^{2}-p^{2}=m^{2}$, with the form $\epsilon^{2}{\tilde f}^{2}(\epsilon) - p^{2}{\tilde g}^{2}(\epsilon)=m^{2}$ where functions ${\tilde f}(\epsilon)$ and ${\tilde g}(\epsilon)$ are commonly known as the rainbow functions. It is worth noting that the rainbow functions are chosen in such a way that they produce, at a low-energy IR limit $\epsilon/M \rightarrow 0$, the standard energy-momentum relation and they are required to satisfy ${\tilde f}(\epsilon) \rightarrow 1$ and ${\tilde g}(\epsilon)\rightarrow 1$ where $M$ is the energy scale that quantum effects of gravity become important. 

Notice that the gravity's rainbow is motivated by modification of usual dispersion relation in the UV limit and captures a modification of the geometry at that limit. Hence, the geometry of the space-time in gravity's rainbow depends on energy of the test particles. Therefore, each test particle of different energy will feel a different geometry of space-time. This displays a family of metrics, namely a rainbow metrics, parametrized by $\epsilon$ to describe the background of the space-time instead of a single metric. In gravity’s rainbow, the modified metric can be expressed as
\ba
g(\epsilon) = \eta^{\mu\nu}{\tilde e}_{\mu}(\epsilon)\otimes{\tilde e}_{\nu}(\epsilon)\,,
\label{act}
\ea
with the energy dependence of the frame field ${\tilde e}_{\mu}(\epsilon)$ can be written in terms of the energy independence
frame field $e_{\mu}$ as ${\tilde e}_{0}(\epsilon)= e_{0}/{\tilde f}(\epsilon)$ and ${\tilde e}_{i}(\epsilon)= e_{i}/{\tilde g}(\epsilon)$ where $i,1,2,3$. In the cosmological viewpoint, the conventional FLRW metric for
the homogeneous and isotropic universe is replaced by a rainbow metric of the form
\ba
ds^{2} = -\frac{1}{{\tilde f}^{2}(\epsilon)}dt^{2} + a(t)^{2}\delta_{ij}dx^{i}dx^{j}\,,
\label{act}
\ea
where $a(t)$ is a scale factor. For convenience, we choose ${\tilde g}(\epsilon)=1$ and only focus on the spatially flat case. As suggested in Ref.\cite{Ling:2006az}, this
formalism can be generalized to study semi-classical
effects of relativistic particles on the background metric during a longtime process. For the very early universe, we consider the evolution of the probe’s energy with cosmic
time, denoted as $\epsilon(t)$. Hence the rainbow functions ${\tilde f}(\epsilon)$ depends on time implicitly through
the energy of particles.

In recent years, gravity's rainbow has attracted a lot of attentions and became the subject of much interest
in the literature. In the context of such gravity, the various physical properties of the black holes are investigated, see e.g. \cite{Feng:2016zsj,Hendi:2016hbe,Feng:2017gms,Hendi:2018sbe,Panahiyan:2018fpb,Dehghani:2018qvn,Upadhyay:2018vfu,Dehghani:2018svw}. In addition, the effects of the rainbow functions have also been discussed in several other scenarios, see for instance \cite{Momeni:2017cvl,Deng:2017umx,Xu:2017luq}. Moreover, the gravity's rainbow was investigated in Gauss-Bonnet gravity \cite{Hendi:2016tiy}, massive gravity \cite{Hendi:2017vgo} and $f(R)$ gravity \cite{Hendi:2016oxk}. More specifically, the gravity’s rainbow has also been used for analyzing the effects of rainbow functions
on the Starobinsky model of $f(R)$ gravity \cite{Chatrabhuti:2015mws}. 

One of the intriguing features of the Starobinsky model \cite{Starobinsky:1980te} is that gravity itself is directly responsible for the inflationary period of the universe without resorting to the introduction of new {\it ad
hoc} scalar fields. The authors of Ref.\cite{Codello:2014sua} studied quantum-induced marginal deformations of the Starobinsky gravitational action of
the form $R^{2(1−\alpha)}$, with $R$ the Ricci scalar and $\alpha$ a positive parameter smaller than one half. The model predicted sizable primordial tensor modes. In the present work, we consider the model proposed by Ref.\cite{Codello:2014sua} in the context of gravity's rainbow.

This paper is organized as follows: In section \ref{ch2}, we revisit the formalism in $f(R)$ theory \cite{Sotiriou:2008rp,DeFelice:2010aj} in the
framework of gravity's rainbow \cite{Chatrabhuti:2015mws}. We then focus on the deformed Starobinsky’s model \cite{Codello:2014sua} in which $f(R)$ takes the form $f(R)\sim R^{2(1-\alpha)}$. We take a short recap of a cosmological linear perturbation in the context of the gravity’s rainbow generated during inflation and calculate the spectral index
of scalar perturbation and the tensor-to-scalar ratio of
the model in section \ref{ch3}. In section \ref{ch4}, we compare the predicted results with Planck data. We conclude our findings in the last section.

\section{$f(R)$ theories with gravity's rainbow effect}
\label{ch2}
As is well known, the modification to general relativity are expected to be plausible in very early universe where possible corrections to Einstein's theory may in principle emerge at high curvature. One of the simplest classes of such modifications is $f(R)$ theories where the Einstein-Hilbert term in the action is replaced by a generic function of the Ricci scalar. We start with the traditionally 4-dimensional action in $f(R)$ gravity \cite{Sotiriou:2008rp,DeFelice:2010aj}. 
\ba
S = \frac{1}{2\kappa^2}\int d^4x \sqrt{-g}f(R) + \int d^4x \sqrt{-g}{\cal L}_{M}(g_{\mu\nu},\Psi_{M})\,,
\label{act}
\ea
where $\kappa^{2}=8\pi G$, $g$ is the determinant of the metric $g_{\mu\nu}$, and the matter field Lagrangian ${\cal L}_{M}$ depends on $g_{\mu\nu}$ and matter fields $\Psi_{M}$. We can derive the field equation by varying the action (\ref{act}) with respect to $g_{\mu\nu}$ to obtain \cite{Sotiriou:2008rp,DeFelice:2010aj}
\ba
F(R)R_{\mu\nu}(g) - \frac{1}{2}f(R)g_{\mu\nu}-\nabla_{\mu}\nabla_{\nu}F(R)+g_{\mu\nu}\Box F(R) = \kappa^{2}T^{(M)}_{\mu\nu}\,,
\label{eom}
\ea
where $F(R)=\partial f(R)/\partial R$ and the operator $\Box$ is defined by $\Box\equiv (1/\sqrt{-g})\partial_{\mu}(\sqrt{-g}g^{\mu\nu}\partial_{\nu})$. Traditionally, the energy-momentum tensor of the matter fields is given by $T^{(M)}_{\mu\nu}=(-2/\sqrt{-g})\delta(\sqrt{-g}{\cal L}_{M})/\delta g^{\mu\nu}$. Here it satisfies the continuity equation such that $\nabla^{\mu}T^{(M)}_{\mu\nu}=0$. Note that the right side of Eq.(\ref{eom}) follows the continuity equation. Using the modified FLRW metric, we can show that the Ricci scalar can be written in terms of the Hubble parameter and a rainbow function as
\ba
R = 6{\tilde f}^{2}\Bigg(2H^{2} + {\dot H} +H\frac{\dot {\tilde f}}{{\tilde f}} \Bigg).\label{eq1r}
\ea
The $(0,0)$-component of Eq.(\ref{eom}) yields the following differential equation:
\ba
3FH^{2} = -3H{\dot F} -{\dot F}\frac{\dot {\tilde f}}{{\tilde f}} + \frac{FR-f(R)}{2{\tilde f}^{2}}+\frac{\kappa^{2}\rho_{M}}{{\tilde f}^{2}}, \label{eq1}
\ea
and the $(i,j)$-component of Eq.(\ref{eom}) reads
\ba
2F\Bigg({\dot H} + H\frac{\dot {\tilde f}}{{\tilde f}}\Bigg) = -{\ddot F}+H{\dot F}-\frac{\kappa^{2}}{{\tilde f}^{2}}\Big(\rho_{M} + P_{M}\Big). \label{eq2r}
\ea
It is worth noting that the above two equations are modified by the rainbow function and these equations can be transformed to the standard ones when setting ${\tilde f}=1$. Let us next consider the deformed Starobinsky model in which $f(R)$ takes the following form \cite{Codello:2014sua}:
\ba
f(R) = R+\frac{R^{2(1-\alpha)}}{6{\tilde M}^{2}}\,,
\label{act1}
\ea
where $1/(6{\tilde M}^{2})=2h m^{4\alpha-2}_{\rm pl}$ and we assume that $\alpha$ is a real parameter with $2|\alpha|<1$ and $h$ is a dimensionless parameter. It is worth noting that the Starobinsky model is recovered when $\alpha=0$. Note that in the standard Starobinsky scenario the $R^{2}$ plays a key role in the very early universe instead of relativistic matter. In the scalar field framework, the Starobinsky theory can be equivalent to the system of one scalar field (an inflaton). It is reasonable if we assume here that the inflaton dominates the very early universe and hence in what follows we can neglect the contributions from matter and radiation, i.e. $\rho_{M}=0$ and $P_{M}=0$.

The combination of Eqs.(\ref{eq1r}), (\ref{eq1}) and (\ref{eq2r}) give us the following system of differential equations:
\ba
{\ddot R}+3H{\dot R}+\frac{4{\dot R}{\dot {\tilde f}}}{3 {\tilde f}} + \frac{{\tilde M}^{2}}{{\tilde f}^{2}}\Bigg(R + \frac{2\alpha}{{\tilde M}^{2}}R^{2(1-\alpha)}\Bigg)= 0\,,
\label{act10}
\ea
and
\ba
&&\frac{1}{{\tilde f}(t)}2 H(t){\dot{\tilde f}}(t)^3+\left(2{\dot{\tilde f}}(t) \left(H(t){\ddot{\tilde f}}(t)+{\dot{\tilde f}}(t) \left(3 {\dot H}(t)+7 H(t)^2\right)\right)+3{\tilde M}^{2} H(t)^2\right)\nonumber\\&&-2{\tilde f}(t) \left(H(t) \left({\dot{\tilde f}}(t) \left(-13 {\dot H}(t)-15 H(t)^2\right)-3 H(t){\ddot{\tilde f}}(t)\right)-{\dot{\tilde f}}(t) {\ddot H}(t)\right)\nonumber\\&&+6{\tilde f}(t)^2 H(t) \left({\ddot H}(t)+5 H(t) {\dot H}(t)+2 H(t)^3\right)+3^{1-2 \alpha } 4^{-\alpha } (2 \alpha -1) \Big(H(t){\dot{\tilde f}}(t)\nonumber\\&&+{\tilde f}(t) \Big({\dot H}(t)+2 H(t)^2\Big)\Big)^2 \left({\tilde f}(t) \left(H(t) {\dot{\tilde f}}(t)+{\tilde f}(t) \left({\dot H}(t)+2 H(t)^2\right)\right)\right)^{-2 \alpha}=0\,\,.\label{act11}
\ea
Note here that when setting $\alpha=0$ our results given in Eqs.(\ref{act10}) and (\ref{act11}) nicely convert to those present in Ref.\cite{Chatrabhuti:2015mws}. In addition, when setting both $\alpha=0$ and ${\tilde f}(t)=1$ our results here reduce to those obtained in the Starobinsky model. Using ${\tilde f}\approx (H/\tilde{M})^{\lambda}$, we obtain
\ba
&&6 (\alpha -1) (2 \alpha -1) (\lambda +1) H \ddot{H} \left(\frac{H}{\tilde{M}}\right)^{2 \lambda }+3 H^2 \Bigg(36^{\alpha } \tilde{M}^2 \left(\left(\frac{H}{\tilde{M}}\right)^{2 \lambda } \left((\lambda +1) \dot{H}+2 H^2\right)\right)^{2 \alpha }\nonumber\\&&+2 (\alpha  (8 \alpha -9)+3) (\lambda +1) \dot{H} \left(\frac{H}{\tilde{M}}\right)^{2 \lambda }\Bigg)+\frac{4 (\alpha -1) (2 \alpha -1) \lambda ^2 (\lambda +1) \dot{H}^3 \left(\frac{H}{\tilde{M}}\right)^{2 \lambda }}{H^2}\nonumber\\&&+(2 \alpha -1) (\lambda +1) ((20 \alpha -17) \lambda +3) \dot{H}^2 \left(\frac{H}{\tilde{M}}\right)^{2 \lambda }\nonumber\\&&+\frac{2 (\alpha -1) (2 \alpha -1) \lambda  (\lambda +1) \dot{H} \ddot{H} \left(\frac{H}{\tilde{M}}\right)^{2 \lambda }}{H}+12 \alpha  H^4 \left(\frac{H}{\tilde{M}}\right)^{2 \lambda }=0. \label{eqH1}
\ea
Since we are only interested in an inflationary solution, it is natural to assume the slow-roll approximation, namely the terms containing $\ddot{H}$ and higher power in $\dot{H}$ can be neglected in this particular regime. Therefore the Eq.(\ref{eqH1}) is reduced to
\ba
\dot{H} \simeq -\frac{\tilde{M}^2 \left(\frac{H}{\tilde{M}}\right)^{-2 \lambda }}{6 (\lambda +1)}-\frac{  \left(\frac{H}{\tilde{M}}\right)^{-2 \lambda }}{6 (\lambda +1)}\Big(2 \tilde{M}^2 \log \Big(12H^2 \Big(\frac{H}{\tilde{M}}\Big)^{2 \lambda }\Big)+4 H^2 \left(\frac{H}{\tilde{M}}\right)^{2 \lambda }+3 \tilde{M}^2\Big)\alpha+{\cal O}(\alpha^{2})\, . \label{eqH2}
\ea
Setting only $\alpha=0$, we obtain the same result given in Ref.\cite{Chatrabhuti:2015mws}. Moreover, setting both $\alpha$ and $\lambda$ to vanish, the result converts to the standard Starobinsky model \cite{Starobinsky:1980te}. Despite the fact that one can numerically solve this equation for the Hubble parameter during inflation, we can find a simple analytical solution to this equation provided the second term on the RHS can be ignored since we are considering a small deviation from the Starobinsky model. Given this further approximation one obtains 
\ba
H &\simeq& H_i - \frac{\tilde{M}^2 \left(\frac{H}{\tilde{M}}\right)^{-2 \lambda }}{6 (\lambda +1)}(t-t_i)\nonumber\\&&\,\,\quad-\frac{  \left(\frac{H}{\tilde{M}}\right)^{-2 \lambda }}{6 (\lambda +1)}\Big(2 \tilde{M}^2 \log \Big(12H^2 \Big(\frac{H}{\tilde{M}}\Big)^{2 \lambda }\Big)+4 H^2 \left(\frac{H}{\tilde{M}}\right)^{2 \lambda }+3 \tilde{M}^2\Big)\alpha(t-t_i)+{\cal O}(\alpha^{2}) \,,\label{Htk}
\ea
and 
\ba
a &\simeq& a_i \exp \Bigg\{H_i(t-t_i) - \frac{\tilde{M}^2 \left(\frac{H}{\tilde{M}}\right)^{-2 \lambda }}{6 (\lambda +1)}(t-t_i)^{2}\nonumber\\&&\quad\quad-\frac{  \left(\frac{H}{\tilde{M}}\right)^{-2 \lambda }}{6 (\lambda +1)}\Big(2 \tilde{M}^2 \log \Big(12H^2 \Big(\frac{H}{\tilde{M}}\Big)^{2 \lambda }\Big)+4 H^2 \left(\frac{H}{\tilde{M}}\right)^{2 \lambda }+3 \tilde{M}^2\Big)\alpha(t-t_i)^{2}+{\cal O}(\alpha^{2})\Bigg\},
\ea
where $H_i$ and $a_i$ are respectively the Hubble parameter and the scale factor at the onset of inflation ($t=t_i$). The slow-roll parameter $\epsilon_1$ is defined by $\epsilon_1 \equiv -\dot{H}/H^2$ which in this case can be estimated to the first order of $\alpha$ as
\ba
\epsilon_1 &\simeq&\frac{H^{-2 (\lambda +1)} \tilde{M}^{2 \lambda +2}}{6 (\lambda +1)}\nonumber\\&&+\frac{H^{-2 (\lambda +1)}}{2 (\lambda +1)} \left(\frac{1}{3} \left(4 H^{2 (\lambda +1)}+\tilde{M}^{2 \lambda +2} \log \left(144 H^{4 (\lambda +1)} \tilde{M}^{4 \lambda }\right)\right)+\tilde{M}^{2 \lambda +2}\right)\alpha\, . \label{epsilon1}
\ea
Note that this parameter is less than unity during inflation ($H^2 \gg \tilde{M}^2$) and we find when setting $\alpha=0$ that $\epsilon_1 \simeq\frac{H^{-2 (\lambda +1)} \tilde{M}^{2 \lambda +2}}{6 (\lambda +1)}$. One can simply determine the end of inflation ($t=t_f$) by the condition $\epsilon(t_f) \simeq 1$, $t_f$ is approximately given by
\ba
t_f &\simeq& t_i +6 (\lambda +1) \frac{H_i^{2 \lambda +1}}{\tilde{M}^{2 \lambda +2}}\nonumber\\&&+\, 6 (\lambda +1) \frac{H_i^{2 \lambda +1}}{\tilde{M}^{4 (\lambda +1)}} \left(-4 H_i^{2 \lambda +2}-\tilde{M}^{2 \lambda +2} \left(\log \left(144 H_i^{2 (\lambda +1)} \tilde{M}^{4 \lambda }\right)+3\right)\right)\alpha \, . \label{tf}
\ea
The number of e-foldings from $t_i$ to $t_f$ is then given by
\ba
N &\equiv& \int^{t_f}_{t_i} Hdt \simeq H_i(t-t_i)- \frac{\tilde{M}^2 \left(\frac{H}{\tilde{M}}\right)^{-2 \lambda }}{6 (\lambda +1)}(t-t_i)^2 \nonumber\\&& -\frac{  \left(\frac{H}{\tilde{M}}\right)^{-2 \lambda }}{6 (\lambda +1)}\Big(2 \tilde{M}^2 \log \Big(12H^2 \Big(\frac{H}{\tilde{M}}\Big)^{2 \lambda }\Big)+4 H^2 \left(\frac{H}{\tilde{M}}\right)^{2 \lambda }+3 \tilde{M}^2\Big)\alpha(t-t_i)^2 \, .
\ea
Using the expressions (\ref{tf}) and (\ref{epsilon1}) the parameter $N$ is thus given by to the first order of $\alpha$: 
\ba
N &\simeq& 3 (\lambda +1) \frac{H^{2 \lambda +2}}{\tilde{M}^{2 \lambda +2}} \nonumber\\&&+ \frac{3 (\lambda +1) H^{2 \lambda +2}}{\tilde{M}^{4 (\lambda +1)}}\Big(-4 H^{2 \lambda +2}-\tilde{M}^{2 \lambda +2} \Big(\log \left(144 H^{4 (\lambda +1)} \tilde{M}^{4 \lambda }\right)+3\Big)\Big)\alpha
= \frac{1}{2\epsilon_1(t_i)} \, . \label{EpN}
\ea
Note that when $\lambda=0=\alpha$ the result is the same as that of the Starobinsky model.
\section{Cosmological Perturbation in Gravity's Rainbow revisited}\label{ch3}
In this section, we will take a short recap of a cosmological linear perturbation in the context of the gravity's rainbow generated during inflation proposed by Ref.\cite{Chatrabhuti:2015mws}. We here begin with a scalar perturbation (since scalar and tensor evolve separately at the linear level) via the following perturbed flat FRW metric taking into account the rainbow effect
\ba
ds^2 = -\frac{1+2\Phi}{\tilde{f}^2(t)}dt^2 + a^2(t)(1-2\Psi)d\vec{x}^2 \, , \label{pertFRW}
\ea
where $\tilde{f}(t)$ denotes the rainbow function. Notice that this perturbed metric has been written in the Newtonian gauge. Let us define a new variable $A \equiv 3(H\Phi+\dot{\Psi})$. With the metric (\ref{pertFRW}) and Eq.(\ref{eom}), we obtain the following system of equations \cite{Chatrabhuti:2015mws}
\ba
-\frac{\nabla^2\Psi}{a^2}+\tilde{f}^2HA &=& -\frac{1}{2F}\left[3\tilde{f}^2\left(H^2 + \dot{H} + \frac{\dot{\tilde{f}}}{\tilde{f}}\right)\delta F + \frac{\nabla^2\delta F}{a^2}-3\tilde{f}^2H\delta\dot{F} \right. \nonumber \\ 
 &+& \left. 3\tilde{f}^2H\dot{F}\Phi + \tilde{f}^2\dot{F}A+\kappa^2\delta\rho_M\right] \, , \label{eomA1} \\
H\Phi+\dot{\Psi}&=&-\frac{1}{2F}(H\delta F+\dot{F}\Phi-\delta\dot{F}) \, ,  \label{eomA2} 
\ea
and 
\ba
\dot{A} + \left(2H+\frac{\dot{\tilde{f}}}{\tilde{f}}\right)A+3\dot{H}\Phi + \frac{\nabla^2\Phi}{a^2\tilde{f}^2}+\frac{3H\Phi\dot{\tilde{f}}}{\tilde{f}} 
= \frac{1}{2F}\left[3\delta\ddot{F}+3\left(H+\frac{\dot{\tilde{f}}}{\tilde{f}}\right)\delta\dot{F} \right. \nonumber \\ -\left. 6H^2\delta F -\frac{\nabla^2\delta F}{a^2\tilde{f}^2} - 3\dot{F}\dot{\Phi}-\dot{F}A - 3\left(H+\frac{\dot{\tilde{f}}}{\tilde{f}}\right)\dot{F}\Phi-6\ddot{F}\Phi+\frac{\kappa^2}{\tilde{f}^2}(3\delta P_M+\delta\rho_M) \right] \, .
\label{eomA3}
\ea
Note that the equations given above can be used to describe evolution of the cosmological scalar perturbations. In what follows, we will solve these equations within the inflationary framework. We first study scalar perturbations generated during inflation and consider not to  take into account the perfect fluid, i.e. $\delta\rho_M =0$ and $\delta P_M=0$. Here we choose the gauge condition $\delta F =0$, so that $\mathcal{R} = \psi = -\Psi$.  Note that the spatial curvature $^{(3)}\mathcal{R}$ on the constant-time hypersurface is related to $\psi$ via the relation $^{(3)}\mathcal{R} = - 4\nabla^2\psi/a^2$. Using $\delta F=0$, we obtain from Eq.(\ref{eomA2}) 
\begin{eqnarray}
 \Phi = \frac{\dot{\mathcal{R}}}{H+\dot{F}/2F} \ , \label{B8}
\end{eqnarray}
and from the equation (\ref{eomA1}), we find
\begin{eqnarray}
 A = -\frac{1}{H+\dot{F}/2F}\left[\frac{\nabla^2\mathcal{R}}{a^2\tilde{f}^2}+\frac{3H\dot{F}\dot{\mathcal{R}}}{2F(H+\dot{F}/2F)}\right] \ . \label{B9}
\end{eqnarray}
Using the background equation (\ref{eq2r}), we find from Eq.(\ref{eomA3})
\begin{align}
 \dot{A}+\left(2H+\frac{\dot{F}}{2F}\right)A+\frac{\dot{\tilde{f}}A}{\tilde{f}}+\frac{3\dot{F}\dot{\Phi}}{2F}+\left[\frac{3\ddot{F}+6H\dot{F}}{2F}+\frac{\nabla^2}{a^2\tilde{f}^2}\right]\Phi+\frac{3\dot{F}}{2F}\frac{\Phi\dot{\tilde{f}}}{\tilde{f}} = 0. \label{B10}
\end{align}
Substituting Eq.(\ref{B8}) and (\ref{B9}) into Eq.(\ref{B10}), we find in Fourier space that the curvature perturbation satisfies the following equation 
\begin{eqnarray}
 \ddot{\mathcal{R}} + \frac{1}{a^3Q_s}\frac{d}{dt}(a^3Q_s)\dot{\mathcal{R}} + \frac{\dot{\tilde{f}}}{\tilde{f}}\dot{\mathcal{R}} + \frac{k^2}{a^2\tilde{f}^2} \mathcal{R}= 0 \ , \label{B13}
\end{eqnarray}
where $k$ is a comoving wave number and $Q_s$ is defined by
\begin{eqnarray}
 Q_s \equiv \frac{3\dot{F}^2}{2\kappa^2F(H+\dot{F}/2F)^2} \ . \label{B12}
\end{eqnarray}
Introducing new variables $z_s = a\sqrt{Q_s}$ and $u = z_s\mathcal{R}$, Eq.(\ref{B13}) can be reduced and then can be expressed as   
\begin{eqnarray}
 u'' + \left(k^2-\frac{z_s''}{z_s}\right)u = 0 \ , \label{B14}
\end{eqnarray}
where a prime denotes a derivative with respect to the new time coordinates $\eta = \int (a\tilde{f})^{-1} dt$. In order to determine the spectrum of curvature perturbations we define slow-roll parameters as
\begin{eqnarray}
 \epsilon_1 \equiv -\frac{\dot{H}}{H^2},  \ \ \epsilon_2 \equiv \frac{\dot{F}}{2HF}, \ \ \epsilon_3 \equiv \frac{\dot{E}}{2HE}\ ,
\end{eqnarray}
where $E \equiv 3\dot{F}^2/2\kappa^2$.   As a result, $Q_s$ can be recast as
\begin{eqnarray}
 Q_s = \frac{E}{FH^2(1+\epsilon_2)^2} \ . \label{B16}
\end{eqnarray}
Here parameters $\epsilon_i$ are assumed to be nearly constant during the inflation and $\tilde{f} \simeq (H/M)^{\lambda}$. These allow us to calculate $\eta$ as $\eta = -1/[(1-(1+\lambda)\epsilon_1)\tilde{f}aH]$. If $\dot{\epsilon_i}\simeq 0$, a term $z_s''/z_s$ satisfies
\begin{eqnarray}
 \frac{z_s''}{z_s} = \frac{\nu^2_{\mathcal{R}} - 1/4}{\eta^2} \ , \label{B18}
\end{eqnarray}
with
\begin{eqnarray}
 \nu_{\mathcal{R}}^2 = \frac{1}{4} + \frac{(1+\epsilon_1 - \epsilon_2+\epsilon_3)(2-\lambda\epsilon_1 -\epsilon_2+\epsilon_3)}{(1-(\lambda+1)\epsilon_1)^2} \ .\label{B19}
\end{eqnarray}
Therefore we find the solution of Eq.(\ref{B14}) written in terms of a linear combination of Hankel functions
\begin{eqnarray}
 u = \frac{\sqrt{\pi|\eta|}}{2}\textmd{e}^{i(1+2\nu_{\mathcal{R}})\pi/4}\left[c_1\textmd{H}_{\nu_{\mathcal{R}}}^{(1)}(k|\eta|)+c_2\textmd{H}_{\nu_{\mathcal{R}}}^{(2)}(k|\eta|)\right] \ , \label{B20}
\end{eqnarray}
where $c_1$, $c_2$ are integration constants and $\textmd{H}_{\nu_{\mathcal{R}}}^{(1)}(k|\eta|)$, $\textmd{H}_{\nu_{\mathcal{R}}}^{(2)}(k|\eta|)$ are the Hankel functions of the first kind and the second kind respectively.
In the asymptotic past $k\eta \rightarrow -\infty$, we find from Eq.(\ref{B20}) $u \rightarrow \textmd{e}^{-ik\eta}/\sqrt{2k}$. This implies $c_1=1$ and $c_2=0$ giving the following solutions
\begin{eqnarray}
 u = \frac{\sqrt{\pi|\eta|}}{2}\textmd{e}^{i(1+2\nu_{\mathcal{R}})\pi/4}\textmd{H}_{\nu_{\mathcal{R}}}^{(1)}(k|\eta|) \ . \label{B21}
\end{eqnarray}
By defining the power spectrum of curvature perturbations
\begin{eqnarray}
 \mathcal{P}_{\mathcal{R}} \equiv \frac{4\pi k^3}{(2\pi)^3}|\mathcal{R}|^2 \ , \label{B22}
\end{eqnarray}
and using Eq.(\ref{B21}) and $u = z_s\mathcal{R}$, we obtain
\begin{eqnarray}
 \mathcal{P}_{\mathcal{R}} = \frac{1}{Q_s}\left[(1-(1+\lambda)\epsilon_1)\frac{\Gamma(\nu_{\mathcal{R}})H}{2\pi\Gamma(3/2)}\left(\frac{H}{M}\right)^\lambda\right]^2\left(\frac{k|\eta|}{2}\right)^{3-2\nu_{\mathcal{R}}} \ , \label{B23}
\end{eqnarray}
where we have used $\textmd{H}_{\nu_{\mathcal{R}}}^{(1)}(k|\eta|) \rightarrow -(i/\pi)\Gamma(\nu_{\mathcal{R}})(k|\eta|/2)^{-\nu_{\mathcal{R}}}$ for $k|\eta| \rightarrow 0$. Since $\mathcal{R}$ is frozen after the Hubble radius crossing, $P_{\mathcal{R}}$ should be evaluated at $k=aH$. Now we define the spectral index $n_{\mathcal{R}}$  as
\begin{eqnarray}
 n_{\mathcal{R}} - 1 = \left.\frac{d\textmd{ln}\mathcal{P}_{\mathcal{R}}}{d\textmd{ln}k}\right|_{k=aH} = 3 - 2\nu_{\mathcal{R}} \ . \label{B25}
\end{eqnarray}
The spectral index can be written in terms of the slow-roll parameters as
\begin{eqnarray}
 n_{\mathcal{R}} - 1 \simeq -2(\lambda+2)\epsilon_1+2\epsilon_2-2\epsilon_3 \ , \label{B27}
\end{eqnarray}
where during the inflationary epoch, we have assumed that $|\epsilon_i | \ll 1$. Notice that the spectrum is nearly scale-invariant when $|\epsilon_i|$ are much smaller than unity, i.e. $n_{\mathcal{R}} \simeq 1$.   Subsequently, the power spectrum of curvature perturbation takes the form
\begin{eqnarray}
 \mathcal{P}_{\mathcal{R}} \approx \frac{1}{Q_s}\left(\frac{H}{2\pi}\right)^2\left(\frac{H}{M}\right)^{2\lambda} \ . \label{B28}
\end{eqnarray}
Note that we obtain the standard result when setting $\lambda=0$ \cite{DeFelice:2010aj}. We next consider the tensor perturbation. In general $h_{ij}$ can be generally written as
\begin{eqnarray}
 h_{ij} = h_{+}e^+_{ij} + h_{\times}e^\times_{ij} \ , \label{C2}
\end{eqnarray}
where $e^+_{ij}$ and $e^\times_{ij}$ are the polarization tensors corresponding to the two polarization states of $h_{ij}$. Let $\vec{k}$ be in the direction along the z-axis, then the non-vanishing components of polarization tensors are $e^+_{xx} = -e^+_{yy} = 1$ and $e^\times_{xy} = e^\times_{yx} = 1$. Without taking into account the scalar and vector perturbation, the perturbed FLRW metric can be written as
\begin{align}
 ds^2 = -\frac{dt^2}{\tilde{f}(\varepsilon)^2} + a^2(t)h_{\times}dxdy + a^2(t)\left[(1+h_{+})dx^2+(1-h_{+})dy^2+dz^2\right] .\label{C1}
\end{align}
Using Eq.(\ref{eom}), we can show that the Fourier components $h_\chi$  satisfy the following equation
\begin{eqnarray}
  \ddot{h}_\chi + \frac{(a^3F)^\cdot}{a^3F}\dot{h}_\chi + \frac{\dot{\tilde{f}}}{\tilde{f}}\dot{h}_\chi + \frac{k^2}{a^2\tilde{f}^2}h_\chi = 0 \ , \label{C5}
\end{eqnarray}
where $\chi$ denotes polarizations $+$ and $\times$. Following a similar procedure to  the case of curvature perturbation, let us introduce the new variables $z_t = a\sqrt{F}$ and $u_\chi = z_t h_\chi /\sqrt{2 \kappa^{2}}$. Therefore Eq. (\ref{C5}) can be written as
\begin{eqnarray}
 u''_\chi + \left(k^2-\frac{z_t''}{z_t}\right)u_\chi = 0 \ . \label{C6}
\end{eqnarray}
Notice that for a massless scalar field $u_\chi$ has dimension of mass. By choosing $\dot{\epsilon}_i=0$, we obtain
\begin{eqnarray}
 \frac{z_t''}{z_t} = \frac{\nu^2_t-1/4}{\eta^2} \ , \label{C7}
\end{eqnarray}
where
\begin{eqnarray}
 \nu^2_t = \frac{1}{4} + \frac{(1+\epsilon_2)(2-(1+\lambda)\epsilon_1+\epsilon_2)}{(1-(1+\lambda)\epsilon_1)^2} \ . \label{C8}
\end{eqnarray}
Similarly the solution to Eq.(\ref{C6}) can be also expressed in terms of a linear combination of Hankel functions. Taking into account polarization
states, the power spectrum of tensor perturbations $P_T$ after the Hubble radius crossing reads  
\begin{align}
 \mathcal{P}_T &= 4\times\frac{2\kappa^{2}}{a^2F}\frac{4\pi k^3}{(2\pi)^3}|u_\chi|^2  \nonumber\\&= \frac{16}{\pi}\left(\frac{H}{M_{P}}\right)^2\frac{1}{F}\left[(1-(1+\lambda)\epsilon_1)\frac{\Gamma(\nu_t)}{\Gamma(3/2)}\left(\frac{H}{M}\right)^\lambda\right]^2\left(\frac{k|\eta|}{2} \right)^{3-2\nu_t},  
 \label{C12}
\end{align}
where we have used $\tilde{f} \simeq (H/M)^{\lambda}$. Therefore $\nu_t$ can be estimated by assuming that the slow-roll parameters are very small during inflation as
\begin{eqnarray}
 \nu_t \simeq \frac{3}{2} + (1+\lambda)\epsilon_1 + \epsilon_2 \ . \label{C13}
\end{eqnarray}
In addition, the spectral index of tensor perturbations is determined via
\begin{eqnarray}
 n_T = \left.\frac{d \textmd{ln}\mathcal{P}_T}{d\textmd{ln}k}\right|_{k=aH} = 3-2\nu_t \simeq -2(1+\lambda)\epsilon_1 - 2\epsilon_2 \ . \label{C16}
\end{eqnarray}
The power spectrum $\mathcal{P}_T$ can also be rewritten as
\begin{eqnarray}
 \mathcal{P}_T \simeq \frac{16}{\pi}\left(\frac{H}{M_{P}}\right)^2\frac{1}{F}\left(\frac{H}{M}\right)^{2\lambda} \ . \label{C17}
\end{eqnarray}
The tensor-to-scalar ratio $r$ can be obtained as
\begin{eqnarray}
 r \equiv \frac{\mathcal{P}_T}{\mathcal{P}_R}  \simeq \frac{64\pi}{M_{P}^2}\frac{Q_s}{F} \ . \label{C20}
\end{eqnarray}
Substituting $Q_s$ from Eq.(\ref{B16}), we therefore obtain
\begin{eqnarray}
 r = 48\epsilon_2^2 \ . \label{C21}
\end{eqnarray}
Let us next examine relations among the slow-roll parameters.  Having assumed that $ |\epsilon_i | \ll 1$ during the inflation and matter field, Eq.(\ref{eq2r}) gives us
\begin{eqnarray}
 \epsilon_2 \simeq -(1+\lambda)\epsilon_1 \ . \label{C23}
\end{eqnarray}
Compared with the Starobinsky model, we have similar form of $ f(R) = R+\frac{R^{2(1-\alpha)}}{6{\tilde M}^{2}}$. Here inflation occurred in the limit $R \gg {\tilde M}^2$ and $|\dot{H}| \ll H^2$. We can approximate $F(R) \simeq (12H^2\tilde{f}^2)^{1-2\alpha}(1-\alpha)/3{\tilde M}^{2}$.  By assuming that $|\epsilon_i| \ll 1$ during the inflation, this leads to
\ba
\epsilon_4 \simeq -(1 + 2\lambda - 4\alpha(\lambda+1))\epsilon_1\,.
\label{ep4}
\ea
Considering Eq.(\ref{B28}), we obtain 
\ba
P_{\cal R}\simeq\frac{144^{\alpha } \left(\epsilon _1+1\right){}^2 {\tilde M}^{2 \alpha +2}}{12 \pi  (1-2 \alpha )^2 (1-\alpha ) (\lambda +1)^2 m^2_{\rm pl} \epsilon ^2_1}.\label{Pr}
\ea
Since $n_\mathcal{R} - 1 \simeq -2(\lambda+2)\epsilon_1+2\epsilon_3-2\epsilon_4$ and $r = 48\epsilon_3^2$, one obtains the spectral index of scalar perturbations and the tensor-to-scalar ratio rewritten in terms of $\epsilon_1$ as follows:
\begin{eqnarray}
 n_\mathcal{R} - 1 \simeq -4(1+2\alpha(\lambda+1))\epsilon_1\  \  \text{and    } \ r \simeq 48(\lambda+1)^2\epsilon_1^2 . \label{C28}
\end{eqnarray}
Let $t_k$ be the time at the Hubble radius crossing ($k=aH$).  From Eq.(\ref{Htk}), as long as the condition $H_i\gg \frac{{\tilde M}^2(t_k-t_i)}{6(1+\lambda)}({\tilde M}/H_i)^{2\lambda}+ {\cal O}(\alpha^{2})$ is satisfied, we can approximate $H(t_k) \simeq H_i$.  The number of $e$-fold from $t=t_k$ to the end of the inflation can be estimated as $N_k \simeq 1/2\epsilon_1(t_k)$. We also find from Eq.(\ref{Pr}) to the leading order of $\alpha$ that
\ba
P_{\cal R}\simeq\frac{{\tilde M}^2 N^2}{3 \pi  (\lambda +1)^2 m^2_{\rm pl}}+\frac{ \left(5 {\tilde M}^2+{\tilde M}^2 \log (144{\tilde M}^{2})\right)N^2}{3 \pi  (\lambda +1)^2 m^2_{\rm pl}}\alpha\,.
\ea
According to the relation (\ref{EpN}), both $n_{\cal R}$ and $r$ can be rewritten in terms of the number of e-foldings as 
\ba
n_\mathcal{R} - 1 = -\frac{2}{N}-\frac{4 \alpha  (\lambda +1)}{N} \label{nsup}
\ea
and 
\ba
r = \frac{12(\lambda+1)^2}{N^2} \,.
\label{r21}
\ea
Notice that the spectral index of scalar perturbations $n_{\cal R}$ does depend on both $\alpha$ and the rainbow parameter, $\lambda$. 

\section{Contact with Observation}
\label{ch4}
\begin{figure}[H]
	\begin{center}
		\includegraphics[width=0.48\linewidth]{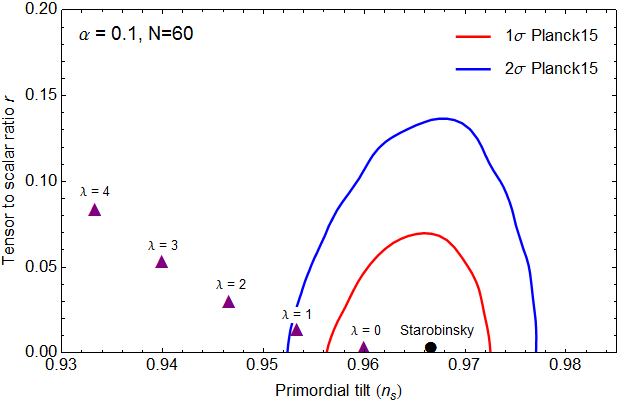}
		\includegraphics[width=0.48\linewidth]{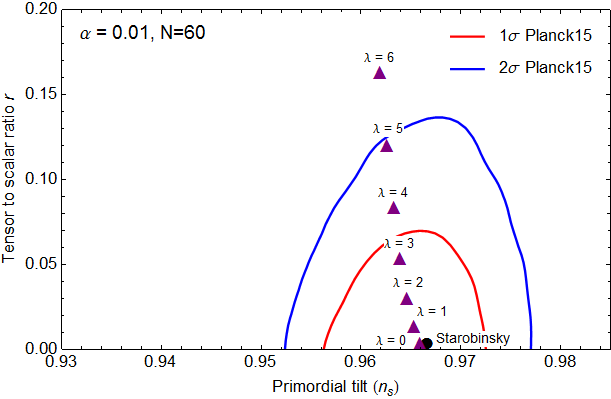}
		\includegraphics[width=0.48\linewidth]{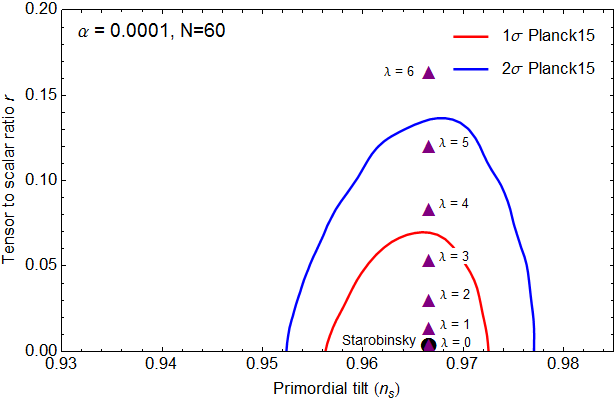}
		\caption{We compare the theoretical predictions in the $(r-n_{s})$ plane for different values of $\lambda$ using $\alpha=0.1,\,N=60$ (uper-left panel); $\alpha=0.01,\,N=60$ (uper-right panel) and $\alpha=0.0001,\,N=60$ (lower panel) with Planck'15
results for TT, TE, EE, +lowP and assuming $\Lambda$CDM+$r$ \cite{Ade:2015lrj}.} \label{sep1}
	\end{center}
\end{figure}

\begin{figure}[H]
	\begin{center}
		\includegraphics[width=0.48\linewidth]{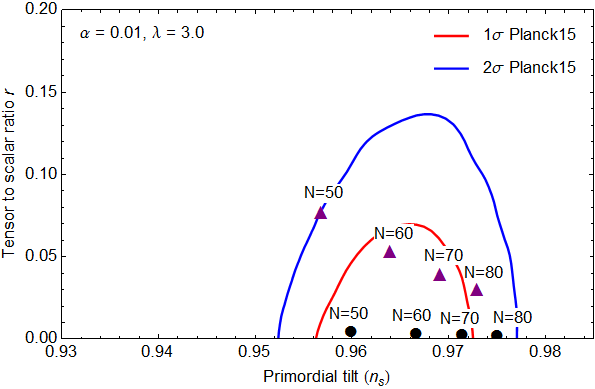}
		\includegraphics[width=0.48\linewidth]{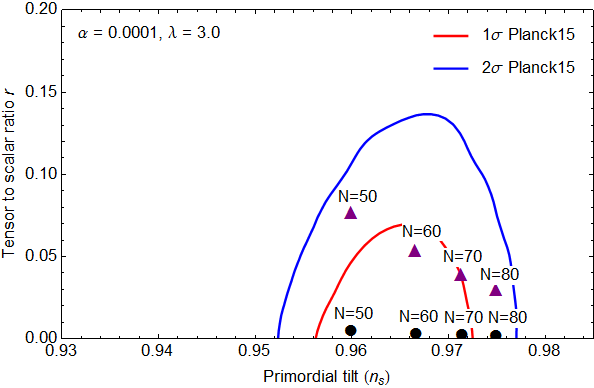}
		\includegraphics[width=0.48\linewidth]{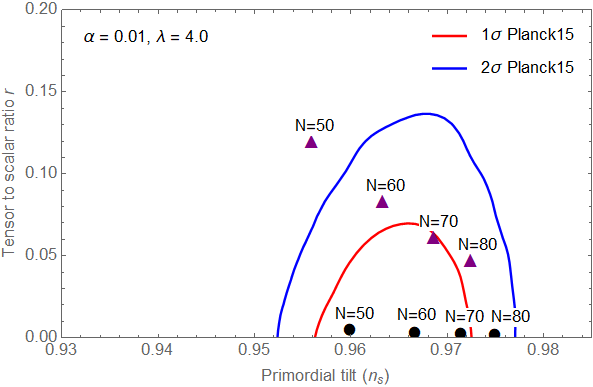}
		\includegraphics[width=0.48\linewidth]{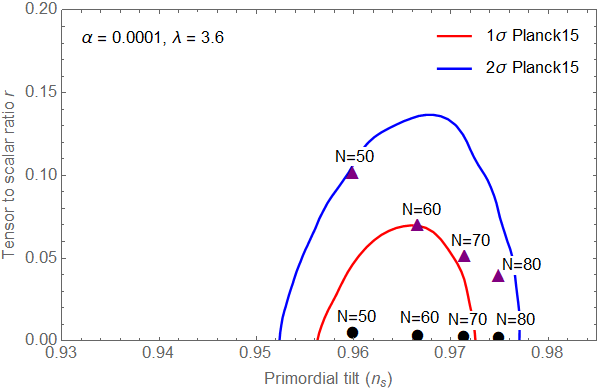}
		\caption{We compare the theoretical predictions in the $(r-n_{s})$ plane for different values of $N$ using $\alpha=0.01,\,\lambda=3.0$ (uper-left panel); $\alpha=0.0001,\,\lambda=3.0$ (uper-right panel); $\alpha=0.01,\,\lambda=5.0$ (lower-left panel) and $\alpha=0.0001,\,\lambda=5.0$ (lower-right panel) with Planck'15
results for TT, TE, EE, +lowP and assuming $\Lambda$CDM+$r$ \cite{Ade:2015lrj}.} \label{sep}
	\end{center}
\end{figure}

\begin{figure}[H]
	\begin{center}
		\includegraphics[width=0.6\linewidth]{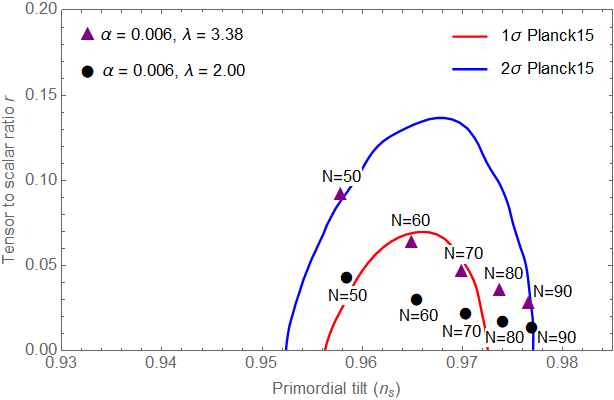}
		\caption{We compare the theoretical predictions in the $(r-n_{s})$ plane for different values of $\lambda$ and $N$ but keep $\alpha$ fixed with Planck'15 results for TT, TE, EE, +lowP and assuming $\Lambda$CDM+$r$ \cite{Ade:2015lrj}.} \label{sep4}
	\end{center}
\end{figure}

In this section, we compare our predicted results with Planck 2015 data. We find from Fig.(\ref{sep1})that the predictions are consistent with the Planck data at two sigma confidence level for $N=60$ only when $\lambda \lesssim 1.00,\,5.00$ and $5.50$ for $\alpha=0.1,\,0.01$ and $0.0001$, respectively. 
We also consider the situation in which the values
of $N$ are arbitrary but keep $\lambda$ and $\alpha$ fixed. From Fig.(\ref{sep}), we observe that in order for the predictions to be satisfied the Planck data at one sigma level a value of $\lambda$ can not be greater than $4.0$ and $3.6$ with $\alpha=0.01$ and $0.0001$, respectively.  
The 2018 recent release of the Planck cosmic microwave background (CMB) anisotropy measurements \cite{Akrami:2018odb} determines the spectral index of scalar perturbations to be $n_{s}=0.9649\pm 0.0042$ at 68\%\,CL and the 95\%\,CL upper limit on the tensor-to-scalar ratio is further
tightened by combining with the BICEP2/Keck Array BK14 data to obtain $r_{0.002} < 0.064$. We use these updated parameters to constrain our model parameters. Let us consider Eq.(\ref{r21}) and then we obtain the upper limit of a parameter $\lambda=\lambda_{*}$ as
\ba
\lambda_{*} < 1.33\times 10^{-2} \Big(5.48\,N-75\Big),
\ea
Interestingly, this upper limit can be used to constrain the value of $\alpha$. Substituting the value of $\lambda_{*}$ into Eq.(\ref{nsup}), we obtain the lower limit of $\alpha=\alpha_{*}$ as
\ba
\alpha_{*} > \frac{2.65\times 10^{-2}}{\left(1.33\times 10^{-2}(5.48 N-75)+1.00\right)}.
\ea
Since a value demanded in most inflationary scenarios is at least $N = 50-60$, we obtain $\lambda_{*}<3.382$ and $\alpha_{*}> 6.06\times 10^{-3}$ for $N=60$. In addition, we compare the theoretical predictions in the $(r-n_{s})$ plane for different values of $\lambda$ and $N$ but keep $\alpha$ fixed with Planck'15 results displayed in Fig.(\ref{sep4}). We find that using $\alpha=0.006$ and $\lambda=3.38$ the predictions in the $(r-n_{s})$ plane lie in the one sigma confidence level only when $N=[60,70]$.
Using parameters of the base $\Lambda$CDM cosmology reported by Planck 2018 \cite{Ade:2015lrj} for $P_{\cal R}$ at the scale $k=0.05$ Mpc$^{-1}$, we
find from Eq.(66) that the mass ${\tilde M}$ is constrained to be
\ba
{\tilde M} \simeq  \sqrt{\frac{3 \pi \kappa _1 (\lambda +1)^2 m_{\rm pl}^2}{N^2}}\Bigg\{1 - \Big(5 + \ln\Big(\frac{432 \pi \kappa _1 (\lambda +1)^2 m_{\rm pl}^2}{N^2}\Big)\Big)\alpha \Bigg\},
\ea
with $\kappa_{1}\approx 2.2065\times 10^{-9}$. In case of very small values of $\alpha$, i.e. $\alpha \ll 0.01$, it becomes
\ba
{\tilde M}&\simeq& \frac{1.41\times 10^{-4}(\lambda +1) m_{\rm pl}}{N}\nonumber\\&\sim & 0.25-1.26\times 10^{14}\,{\rm GeV},
\ea
where the lower value obtained for the reduced Planck
mass of $2.44\times10^{18}$ GeV and the higher one for the
standard one $1.22\times 10^{19}$ GeV and we have used $\lambda=3.38$ and $N=60$. The predicted value of ${\tilde M}$ allows us to further constrain a constant $h$ to obtain
\ba
h \approx \frac{600\, N^{2}}{(1+\lambda)^{2}}.
\ea
Using $N=60,\,\lambda=3.38$ we discover at the scale of ${\tilde M}$ that $h \approx 1.13\times 10^{5}$. However, $h$ is in general scale-dependent. The
explicit computations via heat kernel methods \cite{Avramidi:2000bm} shows that a logarithmic form of $h$ can be induced by leading order quantum fluctuations. The RG improved treatment of $h$ can be found in Ref.\cite{Codello:2014sua}.
\section{Conclusion}
In this work, we studied the deformed Starobinsky model in which the deformations take the form $R^{2(1-\alpha)}$, with $R$ the Ricci scalar and $\alpha$ a positive parameter \cite{Codello:2014sua}. We started by revisiting the formalism in $f(R)$ theory \cite{Sotiriou:2008rp,DeFelice:2010aj} in the framework of gravity's rainbow \cite{Chatrabhuti:2015mws}. We took a short recap of a cosmological linear perturbation in the context of the gravity’s rainbow generated during inflation and calculated the spectral index of scalar perturbation and the tensor-to-scalar ratio predicted by the model. We compared the predicted results with Planck data. With the sizeable number of e-foldings and proper choices of parameters, we discovered that the predictions of the model are in excellent agreement with the Planck analysis. Interestingly, we obtained the upper limit of a rainbow parameter $\lambda < 1.33\times 10^{-2} \Big(5.48\,N-75\Big)$ and found the lower limit of a positive constant $\alpha > 2.65\times 10^{-2}\Big(1.33\times 10^{-2}(5.48 N-75)+1.00\Big)^{-1}$.

Regarding our present work, the study the cosmological dynamics of isotropic and anisotropic universe in $f(R)$ gravity, see e.g.\,\cite{DeFelice:2010aj,Liu:2017edh,Goheer:2007wu} and references therein, via the dynamical system technique can be further studied. Interestingly, the swampland criteria in the deformed Starobinsky model can be worth investigating by following the work done by Ref.\cite{Artymowski:2019vfy}. The reheating process in the present work is worth investigating \cite{Nishizawa:2014zra,Oikonomou:2017bjx}.

\vspace{5mm}
\noindent {\bf Acknowledgments}
The author thanks Vicharit Yingcharoenrat for his early-state collaboration in the present work.



\end{document}